\documentclass[pra,aps,twocolumn,twoside,showpacs]{revtex4}
\usepackage{amsfonts}
\usepackage{amssymb}
\usepackage[dvips]{graphicx}
\usepackage[T1]{fontenc}
\usepackage[latin2]{inputenc}
\usepackage{amsmath}
\usepackage{latexsym}
\usepackage{bbm}

\usepackage{color}

\def\dt{{\rm d}\,}

\newcommand{\mbC}{\mathbb{C}}
\newcommand{\mbF}{\mathbb{F}}
\newcommand{\mbP}{\mathbb{P}}
\newcommand{\mbV}{\mathbb{V}}
\newcommand{\mbW}{\mathbb{W}}
\newcommand{\mbI}{\mathbb{I}}

\def\duzomniejsze{<\kern-.7mm<}
\def\duzowieksze{>\kern-.7mm>}

\def\textbf#1{{\bf #1}}
\def\beq{\begin{equation}}
\def\eeq{\end{equation}}
\def\be{\begin{equation}}
\def\ee{\end{equation}}
\def\ben{\begin{eqnarray}}
\def\een{\end{eqnarray}}
\def\beqa{\begin{eqnarray}}
\def\eeqa{\end{eqnarray}}
\def\eea{\end{array}}
\def\bea{\begin{array}}
\newcommand{\bei}{\begin{itemize}}
\newcommand{\eei}{\end{itemize}}
\newcommand{\bee}{\begin{enumerate}}
\newcommand{\eee}{\end{enumerate}}

\def\hcal{{\cal H}}

\def\1{\openone}

\def\tr{{\rm Tr}}

\def\>{\rangle}
\def\<{\langle}
\def\blacksquare{\vrule height 4pt width 3pt depth2pt}

\def\ot{\otimes}

\def\dt#1{{{\kern -.0mm\rm d}}#1\,}
\def\squareforqed{\hbox{\rlap{$\sqcap$}$\sqcup$}}
\def\qed{\ifmmode\squareforqed\else{\unskip\nobreak\hfil
\penalty50\hskip1em\null\nobreak\hfil\squareforqed
\parfillskip=0pt\finalhyphendemerits=0\endgraf}\fi}

\newtheorem{lemma}{Lemma}
\newtheorem{theorem}[lemma]{Theorem}
\newtheorem{main result}[lemma]{Main result}
\newtheorem{proposition}[lemma]{Proposition}
\newtheorem{definition}[lemma]{Definition}

\newtheorem{fact}[lemma]{Fact}

\newtheorem{corollary}[lemma]{Corollary}

\newtheorem{example}[lemma]{Example}

\newtheorem{remark}[lemma]{Remark}

\newenvironment{mr}[1][Main Result]{\textbf{#1} }

\def\bep{\begin{proposition}}
\def\eep{\end{proposition}}
\def\bel{\begin{lemma}}
\def\eel{\end{lemma}}

\def\bet{\begin{theorem}}
\def\eet{\end{theorem}}
\def\bed{\begin{definition}}
\def\eed{\end{definition}}
\def\bef{\begin{fact}}
\def\eef{\end{fact}}

\begin{document}

\title{Convergence to equilibrium under a random Hamiltonian}

\author{Fernando G.S.L. Brand\~ao${}^1$, Piotr \'Cwikli\'nski$^{2,6,7}$,
Micha\l{} Horodecki$^{3,8}$, Pawe\l{} Horodecki$^{2,8}$, Jaros\l{}aw K. Korbicz${}^4$,
Marek Mozrzymas${}^5$}
\affiliation{
${}^1$ Departamento de F\'isica, Universidade Federal de Minas Gerais, Belo Horizonte, Caixa Postal 702, 30123-970, MG, Brazil \\
${}^2$ Faculty of Applied Physics and Mathematics, Gda\'nsk University of Technology, 80-233 Gda\'nsk, Poland \\
${}^3$ Institute of Theoretical Physics and Astrophysics, University of Gda\'nsk, 80-952 Gda\'nsk, Poland \\
${}^4$ ICFO (Institut de Ci\`{e}ncies Fot\`{o}niques), 08860 Castelldefels (Barcelona), Spain \\
${}^5$ Institute for Theoretical Physics, University of Wroc\l{}aw, 50-204 Wroc\l{}aw, Poland \\
${}^6$ School of Science and Technology, Physics Division, University of Camerino, I-62032 Camerino, Italy \\
${}^7$ Institute for Quantum Information, RWTH Aachen University, D-52056 Aachen, Germany \\
${}^8$ National Quantum Information Centre of Gda\'nsk, 81-824 Sopot, Poland
}
\date{\today}

\begin{abstract}
We analyze equilibration times of subsystems of a larger system
under a random total Hamiltonian, in which the basis of the Hamiltonian
is drawn from the Haar measure. We obtain that the time of equilibration
is of the order of the inverse of the arithmetic average of the Bohr frequencies.
To compute the average over a random basis, we compute the inverse of
a matrix of overlaps of operators which permute four systems. We first obtain results
on such a matrix for a representation of an arbitrary finite group and then apply it to the particular representation of the permutation group under consideration.
\end{abstract}

\maketitle


\section{Introduction}
The phenomenon of convergence to equilibrium despite an underlying deterministic
dynamics was usually justified by referring to subjective lack of knowledge, i.e. by putting probabilities by hand.
However, already in 1929, von Neumann (see \cite{Goldstein2010-vonNeumann}
for an English translation and commentary) put forward an argument for relaxation without referring to an ensemble: For a typical initial pure quantum state, averages of
macroscopic observables will be for most of the time around their equilibrium value.
In this approach, thermalization is implied by statistical properties of quantum states themselves; namely it is due to the fundamental lack of knowledge represented by quantum probability. This "individualist" approach to equilibrium (as phrased in \cite{Goldstein2010-vonNeumann})
has been recently intensively developed; see, e.g., \cite{Mahler2004-thermo,Tasaki1998-thermo,Linden2009-thermo,Goldstein2010-thermo,Reimann2008-thermo, Gong2011}. More broadly, new theoretical and experimental developments on the question of subsystem equilibration in close quantum systems have also been achieved \cite{Rigol2008-thermo,Rigol2009-thermo,Cassidy2011-thermo,Banuls2011-thermo,Cramer2010-thermo,Gogolin2011-thermo,Cramer2008-thermo,Usha2009-thermo,Bloch2008-thermo,Kinoshita2006-thermo,Hofferberth2007-thermo, Short2011-thermo}. However the time of equilibration, a very important aspect of equilibration and thermalization, has not been considered so far. A natural time scale that appears from the analysis of \cite{Linden2009-thermo} is the inverse of the smallest energy gap of the Hamiltonian. However the latter is typically
exponentially small in the size of the system, and thus cannot offer an explanation for the fast nature of thermalization.

In this paper we consider the issue of equilibration time.
As in \cite{Linden2009-thermo} we consider a system $S$ and a bath $B$,
and we are interested in equilibration of the system, given that the bath is sufficiently large.
We evaluate the distance of the state $\rho_{SB}(t)$ of the system and the bath,
evolving according to a random Hamiltonian, from the state $\omega_{SB}$
which is obtained by removing the blocks of $\rho_{SB}(0)$ which are off-diagonal
with respect to the Hamiltonian spectral decomposition.

Our main result amounts to showing that if we choose
the eigenbasis of the Hamiltonian randomly according to the Haar measure,
then the equilibration time depends on the (weighted) average
distance between the energies of the Hamiltonian rather than on the worst case gap.

Computing the  average over the random choice of the eigenbasis of the Hamiltonian,
is reduced to evaluating averages of the sort $\tr \biggl[U^{\ot 4} X (U^{\ot 4})^\dagger Y \biggr]$
over the Haar distributed unitary transformations $U$, with $X,Y$ being some operators. This leads us to
a general problem  of inverting a matrix $M_{gh}= \chi (g^{-1} h)$, where $g,h$ are elements of a finite group $G$
and $\chi$ is a character of some given representation of the group. It turns out that such a matrix
enjoys certain nice properties,
which allow us to obtain the inverse in the case of interest (i.e., for $G=S_4$). We also present
some other properties of the above matrix.

The main results of the work can be summarized in the following statement (see Sec. \ref{sec:eig}):

\begin{mr} {\it For an ensemble of random Hamiltonians with eigenbases distributed according to the
corresponding Haar measure and a not too big level degeneracy [see Eq.~(\ref{low_degen})], the following holds:

1) For an additionally not too big energy gap degeneracy [see Eq.~(\ref{low_degen_gap})], the convergence to equilibrium 
happens at the time scale
of the order of the (weighted) average inverse energy gap $|E_i-E_j|^{-1}$ and the (weighted) average inverse 
second gap $|E_i-E_j-E_k+E_l|^{-1}$ [see Eqs.~(\ref{delta},\ref{deltadelta})];

2) For a simplified model with the energies distributed according to independent
Gaussian measures with variance of the order of $\log d$, where $d$ is the total dimension of the 
system and bath, the convergence to equilibrium happens at the time scale
of the order of $1/(\log d)$.}
\label{main}
\end{mr}

In what follows we prove the above results in the following steps: In Sec.~\ref{sec:av}
we calculate the Haar measure average of the distance from the equilibrium state
over a random basis of a Hamiltonian.
Then in Sec.~\ref{sec:eig} we investigate the dependence of the equilibration time on
the eigenvalues of a random Hamiltonian and derive our main results.
We conclude with some general remarks and connections to other works.
In the appendices we present
the group theoretical machinery needed to perform the Haar measure average
from Sec.~\ref{sec:av}.


\section{Averaging over a random choice of the eigenbasis}
\label{sec:av}
\begin{figure}[h!]
\begin{center}
\includegraphics[width=0.45\textwidth, height=0.40\textheight]{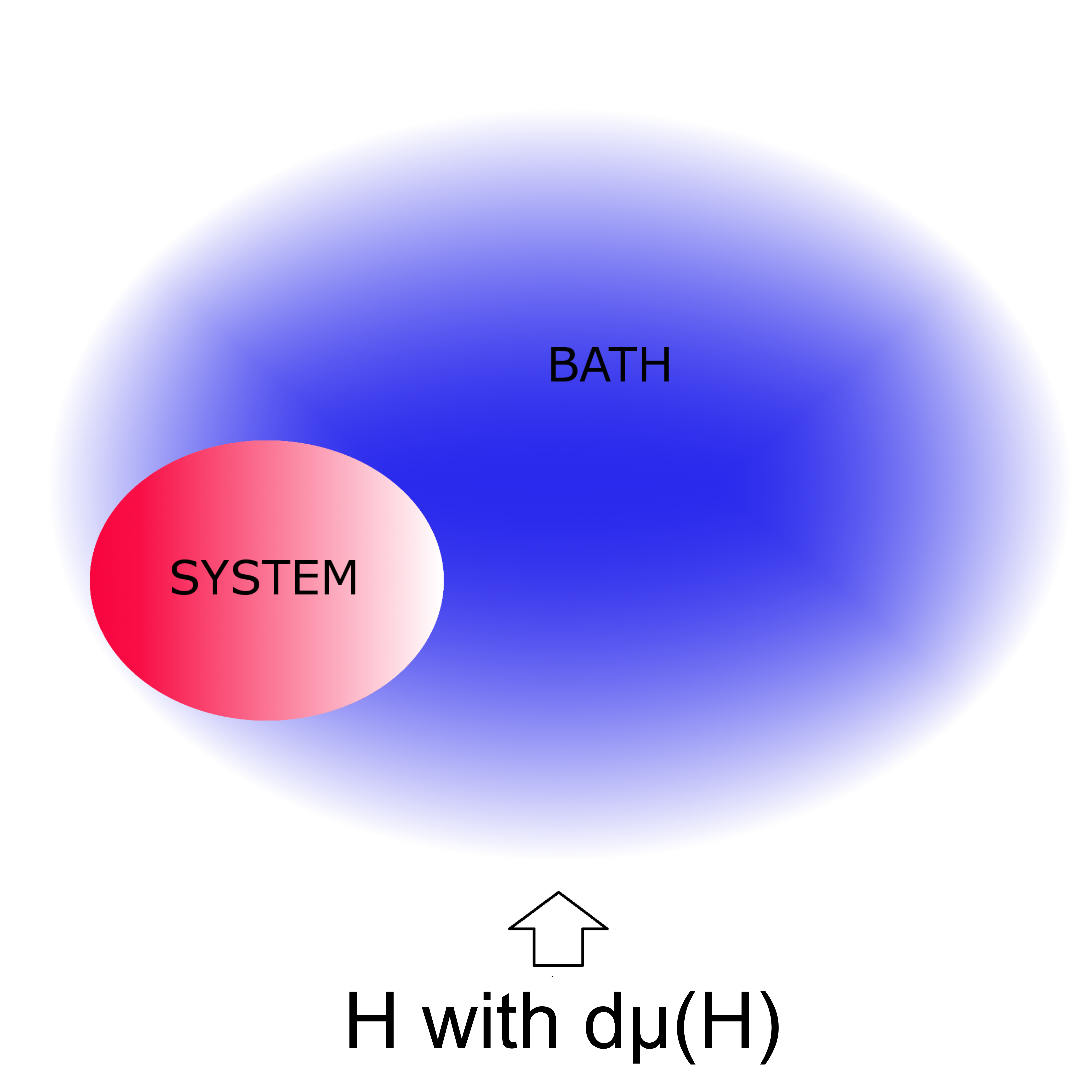}
\end{center}
\caption{The composite system consisting of a system and a bath, governed by a random Hamiltonian with the eigenbasis drawn according to the Haar measure.}
\label{sysba}
\end{figure}

Let us introduce some notation. We consider two systems $S$ (the system) and $B$ (the bath), with the latter playing the role of a heat bath (see Fig. \ref{sysba}). 
The composite system $SB$ is in an arbitrary initial state $\rho_{SB}(0)=|\psi\>_{SB} \<\psi|$.
Since we shall consider random Hamiltonians, whose eigenbases are chosen according to the Haar distribution,
we can equally well take a standard product initial state:
$|\psi_{SB}\>=|0\>_S|0\>_B$. We now consider the evolved state $\rho_{SB}(t)$ given
by
\be
\rho_{SB}(t)=e^{-iHt} \rho_{SB}(0)e^{iHt},
\label{eq:rhot}
\ee
where $H$ is the total Hamiltonian of the system and the bath. We also define a state $\omega_{SB}$ as
\be
\omega_{SB}=\sum_i P_i \rho_{SB}(0)P_i,
\ee
where $P_i$ are eigenprojectors of the Hamiltonian
\be
H=\sum_i E_i P_i.
\ee
We set
\be
H=U H_0 U^\dagger, \quad P_i=U P_i^0 U^\dagger \label{H0},
\ee
where $H_0$ denotes the diagonal Hamiltonian with the elements being (possibly degenerated)
eigenenergies, connected to a given eigenprojector.
We assume that the probability measure over random Hamiltonians splits into two parts
\be
{\rm d}\mu (H)={\rm d}U {\rm d}\mu_2 (H_0),
\ee
where ${\rm d}U$ is the Haar measure, while $\mu_2$ is some distribution over the
eigenenergies (such a separation holds, e.g., for Gaussian unitary ensembles).
Therefore for the averages we have that $\<\ldots\>_H=\<(\<\ldots\>_U)\>_{H_0}$
Let us also introduce the following notation:
$W=e^{iHt}$, $W_0=e^{it H_0}$, $\mbW=W_0\ot W_0^\dagger$, and $\mbP=\sum_i P_i^0\ot P_i^0$.
By  ${\mathbb V}_{X_1:X_2}$ we will
denote the operator which swaps the systems $X_1$ and $X_2$.

We consider the  distance between the reduced state $\rho_S(t)=\tr_B \rho_{BS}(t)$ and
the corresponding reduced equilibrium
state $\omega_S=\tr_B(\omega_{SB})$, induced by the Hilbert-Schmidt norm $ ||A||_2=\sqrt{\tr(A^\dagger A)} $.
Our main goal is to average it over random Hamiltonians. In this section
we will compute the average over the Haar measure.
To this end we will need the following
\begin{proposition}
The following relation holds
\be
\tr(\rho_S(t)-\omega_S)^2= \tr \biggl[Y U^{\ot 4} X 	{U^\dagger}^{\ot 4} \biggr],
\label{eq:dist}
\ee
where
\begin{eqnarray}
& &X=(\mbW-\mbP)_{13}\ot (\mbW^\dagger-\mbP)_{24}, \label{eq:X}\\
& &Y=\mbV_{12:34} (\sigma_{1} \otimes \sigma_{2}\ot \mbF_{34}),\label{eq:Y}
\end{eqnarray}
with $\sigma=\rho_{SB}(0)$, $\mathbb{F}_{34}=\mathbb{V}_{S_3:S_4}\ot \mbI_{B_3:B_4}$, and
the label $i=1,2,3,4,$ denoting a copy of the composite system $S_iB_i$.
\label{prop:dist-U4}
\end{proposition}
The proof is based on the following easy-to-check relation,
coming from the basic properties of the swap operator
and true for any two systems $1$ and $2$, and for arbitrary operators $A_1$, $B_2$, $C_{12}$ and $D_{12}$
\cite{Werner-quantumstates}:
\ben
&&\tr[(C_{12} A_1 \otimes B_2) (D_{12} A^{\dag}_1 \otimes B^{\dag}_2)] = \nonumber \\
&&\tr[\mathbb{V}_{12:34} (C_{12} \otimes D_{34}) (A_1 \otimes B_2 \otimes A^{\dag}_{3} \otimes B^{\dag}_{4})]\label{fac:ABCD}
\een
(here $3$ and $4$ are auxiliary systems, isomorphic to $1$ and $2$, respectively).
The details of the proof are given in Appendix A.

Using Proposition~\ref{prop:dist-U4} we now prove the main result of this section:
\begin{theorem} \label{maontheo}
The Haar measure average of the distance (\ref{eq:dist}) is given by
\be
\<||\rho_{S}(t) - \omega_{S}||_{2}^2\>_U = \frac{|\eta|^2}{d^2}\frac{1}{d_S}+
\left( \frac{|\xi|^2}{d^2} - \frac{\gamma}{d^2} \right)^2 + O(\frac{1}{d_B}),
\label{eq:main}
\ee
where
\ben
&\xi = \tr W_0 = \sum_j d_j e^{i E_j t}, \ \eta = \tr W^2_0 = \sum_j d_j e^{2 i E_j t}, \nonumber \\
&\gamma = \sum_j d_j^2, \ d = d_Sd_B, \ \sum_j d_j = d,
\een
index $j=1,\dots,N$ enumerates the non-degenerate energy levels of the random Hamiltonian $H$,
$d_j$'s are (fixed) energy degeneracies, and $\<\cdot\>_U$ denotes the average
according to the corresponding Haar measure.
\label{thm:main}
\end{theorem}
{\bf Remark.} Note that in Refs.~\cite{Vinayak2011-thermo, Masanes2011-thermo, Cramer2011-thermo}, similar bounds were obtained for the expected distance of $\rho_{S}(t)$ to the equilibrium state $\omega_S$.

Before we proceed with the proof, we briefly note that in the non-degenerate case,
i.e. when all $d_i=1$, Eq.~(\ref{eq:main}) reduces to
\be
\<||\rho_{S}(t) - \omega_{S}||_{2}^2\>_U = \frac{|\eta|^2}{d^2}\frac{1}{d_S}+
\frac{|\xi|^4}{d^4} + O(\frac{1}{d_B}).
\ee
{\bf Proof of the Theorem \ref{thm:main}.}
Thanks to Proposition~\ref{prop:dist-U4}, calculation of the Haar measure average of the distance
$\int d U ||\rho_{S}(t) - \omega_{S}||_{2}^2$ is reduced to a computation of a trace
$\tr[Y \tau_4 (X)]$, where $\tau_4(\cdot)=\int d U U^{\ot 4}(\cdot) {U^{\ot 4}}^\dagger$ is
a twirling operator and $X,Y$ are given by \eqref{eq:X} and \eqref{eq:Y} respectively:
\be
\int d U ||\rho_{S}(t) - \omega_{S}||_{2}^2=\tr[Y \tau_4 (X)].\label{trace}
\ee
Such traces can be dealt with in a systematic manner using group theory, in this case
the representation theory of the permutation group of four elements $S_4$ (c.f.
Appendix B), which greatly simplifies the calculations.

Our main tool will be Proposition~\ref{fact:twirl} from Appendix B.
To apply it, we first express the operators $X$ and $Y$ in terms of product operators:
\be
X=C_1 - C_2 - C_3 + C_4,
\ee
where
\ben
&&C_1= W_0 \ot W_0^\dagger \ot W_0 \ot W_0^\dagger,\nonumber\\
&&C_2= \sum_i W_0 \ot W_0^\dagger \ot P_i^0\ot P_i^0,\nonumber\\
&&C_3= \sum_i P_i^0\ot P_i^0\ot W_0 \ot W_0^\dagger,\nonumber\\
&&C_4= \sum_{ij} P_i^0\ot P_i^0\ot P_j^0\ot P_j^0,
\een
and 
\be
Y=\sum_{ij} \sigma_1 \ot \sigma_2 \ot A^{ij}_3 \ot A^{ji}_4,
\ee
where $A^{ij} = |i\>_S \<j| \ot \mbI_B$ and $|i\>_S$, $|j\>_S$ form an orthonormal basis of the system. Note that in each case we have ordered the systems in the following way $(3, 4, 1, 2)$.

For operators $C_k$, $k=1,\dots, 4$, and $Y$ given above,
we define vectors $\vec c^{(k)}$ by $c^{(k)}_\pi=\tr C_k \mbV_{\pi^{-1}}$  and $\vec a$
by $a_\pi=\tr (Y \mbV_{\pi^{-1}})$, where $\pi$ runs through the elements of the permutation group
$S_4$. 
In order to compute the above vectors, we decompose a given permutation $\pi$ into cycles,
so that for product operators the vector components break into products of separate terms, associated with the cycles.
For a single cycle we then use Proposition~\ref{fact:cycle} from Appendix B. We obtain
\ben
\vec c_1 = \nonumber &&(| \xi |^{4}, | \xi |^{2}d, | \xi |^{2}d, | \xi |^{2}, | \xi |^{2}, \overline{\eta}\xi^{2}, | \xi |^{2}d, d^{2}, | \xi |^{2}, d, d, | \xi |^{2}, \nonumber \\
&& | \xi |^{2}, d, \eta \overline\xi^{2}, | \xi |^{2}, | \eta |^{2}, d, d, | \xi |^{2}, | \xi |^{2}, | \xi |^{2}d, d, d^{2}), \nonumber \\
\vec c_2 = \nonumber &&(\gamma|\xi|^{2}, d, p \xi^{*}, \gamma, \gamma, p^{*}\xi, d|\xi|^{2}, d^2, |\xi|^{2}, d, d, |\xi|^{2}, \nonumber \\
&& |\xi|^{2}, d, p \xi^{*}, \gamma, \gamma, d, d, | \xi |^{2}, \gamma, p^{*}\xi, d, \gamma),
\nonumber \\
\vec c_3 = \nonumber &&(\gamma|\xi|^{2},d,p\xi^{*},\gamma,\gamma,p^{*}\xi,d|\xi|^{2},d^{2},|\xi|^{2},d,d,|\xi|^{2},|\xi|^{2},d, \nonumber \\
&&p \xi^{*}, \gamma,\gamma,d,d,|\xi|^{2},\gamma,p^{*}\xi,d,\gamma), \nonumber \\
\vec c_4=  \nonumber &&(\gamma^{2},\gamma d,\iota,\gamma,\gamma, \iota,\gamma d,d^{2},\gamma,d,d,\gamma, \nonumber \\
&&\gamma,d, \iota,\gamma,\gamma,d,d,\gamma,\gamma, \iota,d,\gamma),
\nonumber \\
\vec a = \nonumber &&(1, 1, d_S, d_S, d_B, d_B, 1, 1, d_S, d_S, d_B, d_B,   \nonumber \\
&& d_B, d_B, d_B, d_B, d d_B, d d_B, d_S, d_S, d_S, d_S, d d_S, d d_S).\label{a'}
\een
Here, $\xi_{i}= \tr (P_i W_0)= d_i e^{i E_i t}$,  $p = \sum_i d_i \xi_i$, $\gamma =\sum_i d_{i}^{2}$, and $\iota = \sum_i d_i^3$.

Now, we proceed to compute the matrix $M^{-1}$ from Proposition~\ref{fact:twirl} from Appendix B.
We refer to Sec.~\ref{Ap:B} of Appendix B, where we consider the general properties of the matrix
$M$ defined for representation of any group. In our case,
for $d \geq 4$ the matrix is invertible, and its inverse is given by \eqref{eq:inverse-M}.
We can now use formulas (\eqref{trace}-\eqref{a'}) and \eqref{eq:inverse-M} to finally obtain
\ben
&& \<||\rho_{S}(t) - \omega_{S}||_{2}^2\>_U = \nonumber
-\frac{1}{d^2 (2+d) (-3-d+3 d^2+d^3)} \times \nonumber \\
&&(4 d+4 d^2+2 d^3-4 d d_B+4 d^3 d_B+d^4 d_B - \nonumber \\
&&|\xi|^4 (2+d) (1+d-d_B-d_S)-4 d d_S-2 d^3 d_S - \nonumber \\
&&4 d^4 d_S-d^5 d_S-b (2+(-2+2 d+4 d^2+d^3) d_B - \nonumber \\
&&(2+4 d+d^2) d_S)+4 \gamma-4 d \gamma-6 d^2 \gamma-2 d^3 \gamma - \nonumber \\
&&4 d_B \gamma+2 d d_B \gamma+5 d^2 d_B \gamma+d^3 d_B \gamma-4 d_S \gamma+ \nonumber \\
&&2 d d_S \gamma+5 d^2 d_S \gamma+d^3 d_S \gamma-2 \gamma^2-3 d \gamma^2 - \nonumber \\
&&d^2 \gamma^2+2 d_B \gamma^2+d d_B \gamma^2+2 d_S \gamma^2+d d_S \gamma^2+ \nonumber \\
&&2 |\xi|^2 (1+d-d_B-d_S) (d+2 \gamma+d \gamma)+4  \iota+ \nonumber \\
&&4 d  \iota-4 d_B  \iota-4 d_S  \iota+\overline{\xi}^2 \eta+d \overline{\xi}^2 \eta-d_B \overline{\xi}^2 \eta - \nonumber \\
&&d_S \overline{\xi}^2 \eta+\xi^2 \overline{\eta}+d \xi^2 \overline{\eta}-d_B \xi^2 \overline{\eta}-
d_S \xi^2 \overline{\eta}-4 p^{*} \xi - \nonumber \\
&&4 d p^{*} \xi +4 d_B p^{*} \xi +4 d_S p^{*} \xi -4
p \xi^{*}-4 d p \xi^{*}+ \nonumber \\
&&4 d_B p \xi^{*}+4 d_S p \xi^{*} )
\een
One then finds that up to the order of $1/d_B$ and a constant factor,
this gives the right hand side of \eqref{eq:main}. $\blacksquare$

We finish this section with two remarks.

First, we note that using ideas of measure concentration \cite{Chatterjee2007-cm} it is easy to show that Theorem \ref{maontheo}
can be extended to say that the vast majority of unitaries $U$ will have the distance $||\rho_{S}(t) - \omega_{S}||_{2}^2$
close to the average and hence the corresponding Hamiltonian will equilibrate quickly.
Moreover, we can pass to the trace norm by using the norm inequality \cite{Horn1985-book-mat} $||A||_1 \leq \sqrt{D}||A ||_2$,
valid for any operator $A$ acting on $\mbC^D$, which adds a factor of $d_S$ (recall that
we consider $d_B \gg d_S$).

Second, let us recall Levy's lemma \cite{Popescu-thermo}
\begin{theorem}
\label{lemma:levy}
For a Lipschitz continuous function $f$, it holds
\be
\Pr_{U \sim \mu_{\text{Haar}}} \left( |f(U) - \langle f\rangle_U| \geq \delta  \right) \leq C e^{-c d \delta^{2}},
\label{levy}
\ee
where $C, c$ are constants and $d$ is the dimension of the total system.
\end{theorem}
We apply Levy's lemma~\ref{lemma:levy} to the average of Theorem~\ref{maontheo}, putting $\delta = d^{-\frac{1}{3}}$. After passing to the trace norm, we then obtain that with a high probability (according to the Haar measure) the following holds:
\ben
&&||\rho_S(t) - \omega_S||_1 \leq \nonumber \\
&&c \left\{\frac{|\eta|}{d}+
\sqrt{d_S}\frac{|\xi|^2}{d^2} - \sqrt{d_S} \frac{\gamma}{d^2} + O(\frac{d_S}{d_B}) + \sqrt{\frac{d_S}{d^{1/3}}}\right\},
\een
where $c$ is an absolute constant and the other notation is as in Theorem \ref{maontheo}.

\section{Average over time/energies}
\label{sec:eig}
In the previous section we have obtained expression (\ref{eq:main}), which depends only on eigenvalues.
Here we will consider the average over time, for a fixed spectrum, and also the average over the Gaussian distributed spectrum.

Using Eq.~(\ref{eq:main}) and averaging over a fixed time interval $[0, T]$, we find
\begin{eqnarray}
&& \frac{1}{T} \int_{0}^{T} dt \<||\rho_{S}(t) - \omega_{S}||_{2}^2\>_U= \nonumber \\ 
&=& \frac{\gamma}{d^2d_S}+\frac{2\gamma^2}{d^4}\nonumber \\ 
&+& \frac{1}{Td^{2}}\sum_{j>k} \left(\frac{d_jd_k}{d_S}+\frac{d_j^2d_k^2}{d^2}\right)
\frac{\sin\big[2T(E_j - E_k)\big]}{(E_j - E_k)} \nonumber \\ 
&+& \frac{2}{Td^{4}}\sum_{j>k}\sum_{\substack{r>s\\ (rs)\ne (jk)}} d_jd_kd_rd_s\Bigg\{\frac{\sin\big[T(E_j-E_k+E_r-E_s)\big]}{(E_j-E_k+E_r-E_s)} \nonumber \\
&+& \frac{\sin\big[T(E_j-E_k-E_r+E_s)\big]}{(E_j-E_k-E_r+E_s)}\Bigg\}. \label{T-average1}
\end{eqnarray}

From the above it is clear that one has to take into account not only the level degeneracies $d_j$, 
but also gap degeneracies. We order the 
energies $E_1<E_2<\dots$, so that for $j>k$, $\Delta_{jk}\equiv(E_j-E_k)>0$ and introduce the following
gap degeneracy related constants 
\begin{equation}
\gamma_{jk}\equiv\sum_{\substack{r>s\\ (rs)\ne (jk)\\ \Delta_{rs}=\Delta_{jk}}}d_rd_s.
\end{equation}
Then the average (\ref{T-average1}) can be rewritten as:

\begin{eqnarray}
&& \frac{1}{T} \int_{0}^{T} dt \<||\rho_{S}(t) - \omega_{S}||_{2}^2\>_U= \nonumber \\ 
&=& \frac{\gamma}{d^2d_S}+\frac{2\gamma^2}{d^4}+2\sum_{j>k} \frac{\gamma_{jk}d_jd_k}{d^4}\nonumber \\ 
&+& \frac{1}{Td^{2}}\sum_{j>k} \left(\frac{d_jd_k}{d_S}+\frac{d_j^2d_k^2}{d^2}+\frac{\gamma_{jk}d_jd_k}{d^2}\right)
\frac{\sin(2T\Delta_{jk})}{\Delta_{jk}} \nonumber \\ 
&+& \frac{2}{Td^{4}}\sum_{j>k}\sum_{\substack{r>s\\ (rs)\ne (jk)\\ \Delta_{rs}\ne\Delta_{jk}}} d_jd_kd_rd_s
\Bigg\{\frac{\sin\big[(T(\Delta_{jk}+\Delta_{rs})\big]}{\Delta_{jk}+\Delta_{rs}} \nonumber \\
&+& \frac{\sin\big[T(\Delta_{jk}-\Delta_{rs})\big]}{\Delta_{jk}-\Delta_{rs}}\Bigg\}. \label{T-average2}
\end{eqnarray}

From (\ref{T-average2}) it follows that the system will have a chance to equilibrate if both the energy and the energy gap 
degeneracies are not too big, i.e. when:
\begin{eqnarray}
&& \frac{\gamma}{d^2}=O(\frac{1}{d}),\label{low_degen}\\
&& \frac{1}{d^4}\sum_{j>k}\gamma_{jk}d_jd_k=O(\frac{1}{d}).\label{low_degen_gap}
\end{eqnarray}
Assuming the above, we obtain the following upper bound (using the trivial estimates $|\sin x|\leq 1$ and $1/(\Delta_{jk}+\Delta_{rs})
\leq 1/|\Delta_{jk}-\Delta_{rs}|$; by our convention all $\Delta_{jk}>0$):
\begin{eqnarray}
&& \frac{1}{T} \int_{0}^{T} dt \<||\rho_{S}(t) - \omega_{S}||_{2}^2\>_U \leq\nonumber\\
&\leq& \frac{1}{T}\Big\{\sum_{j>k}\frac{d_jd_k}{d^2d_S}\frac{1}{\Delta_{jk}}\nonumber\\
&+& 4\sum_{j>k}\sum_{\substack{r>s\\ (rs)\ne (jk)\\ \Delta_{rs}\ne\Delta_{jk}}} \frac{d_jd_kd_rd_s}{d^4}
\frac{1}{|\Delta_{jk}-\Delta_{rs}|}\Big\}+O(\frac{1}{d}).
\end{eqnarray}
Thus for $T$ greater than the bigger of the weighted averages,
\be
T \gg \max\left\{\frac{1}{d_S}\<\Delta_{jk}^{-1}\>,\<|\Delta_{jk}-\Delta_{rs}|^{-1}\>\right\},
\ee
where
\ben
&&\<\Delta_{jk}^{-1}\>\equiv\frac{1}{d^2} 
\sum_{j>k} \frac{d_jd_k}{\Delta_{jk}} \label{delta}\\
&&\<|\Delta_{jk}-\Delta_{rs}|^{-1}\>\equiv \nonumber \\
&& \equiv\frac{1}{d^4}\sum_{j>k}\sum_{\substack{r>s\\ (rs)\ne (jk)\\ \Delta_{rs}\ne\Delta_{jk}}} 
\frac{d_jd_kd_rd_s}{|\Delta_{jk}-\Delta_{rs}|}\label{deltadelta}
\een
the state of the subsystem is close to the asymptotic state $\omega_{S}$.
This proves first part of our main result, stated in the Introduction.

%

Next, we proceed to calculate the average of Eq.~(\ref{eq:main}) over the eigenenergies $E_i$.
For the purpose of this work, we will only consider a simplified situation
(c.f. Ref.~\cite{Masanes2011-thermo} for a more general albeit asymptotic result),
where the probability measure over $E_i$ is: i) product in the energies (we neglect energy repulsion); ii) Gaussian, i.e. we consider the following distribution
\be
\varrho(E_1, \ldots E_N)=\varrho_0(E_1) \ldots  \varrho_0(E_N),
\label{eq:indep}
\ee
where $N$ is the number of non-degenerate energy levels and
\be \label{form}
\varrho_0(E_j)\equiv \frac{1}{\sqrt{2\pi} \sigma} e^{-\frac{E_j^2}{2\sigma^2}}.
\ee
As the energy scale $\sigma$  for the purpose of this work we choose $\sigma=\log d$. 

The latter choice is motivated by the following reasoning.
We may view a $d$-dimensional space as composed of  $\log(d)$ abstract elementary systems (qubits).
Since we want the energy to be extensive, it should then scale as $\log(d)$. Assuming the worse case scenario
that the uncertainty in the energy is of the order of the energy itself leads to $\sigma=\log(d)$ and we obtain:

\begin{theorem}
For an ensemble of random Hamiltonians, satisfying (\ref{low_degen}) and
described by the Haar measure
and the energy distribution \eqref{eq:indep} we have
\begin{eqnarray}
&& \frac{N^2-N}{d_S}e^{-4t^2 (\log d)^2} + O\bigl(\frac{1}{d}\bigr) \lesssim \<||\rho_S(t)-\omega_S||^{2}_{2}\>_H \nonumber \\
&&\lesssim (N^2-N)^2e^{- t^2 (\log d)^2} + O\bigl(\frac{1}{d}\bigr).\label{thm_gauss}
\end{eqnarray}
In the above we used the following average:
\ben
&&\< f(t, E_1 \ldots E_N) \>_{H_0} \equiv \< f(t, E_1 \ldots E_N) \>_{ \left\{ E_k \right\} } \nonumber \\
&&\equiv \int f(t, E_1 \ldots E_N) \varrho_0(E_1) \ldots \varrho_0(E_N) dE_1 \ldots dE_N, \een
where $\varrho_0(E_j)$ are of the form \eqref{form}.
\end{theorem}

This Theorem proves our second main result, stated in point 2) in the Introduction.
As mentioned there, it shows that, under the above conditions,
the time of convergence of the state $\rho_S(t)$ to equilibrium
scales roughly as the inverse of $\log d$, i.e. as the inverse of the volume of the
total system (in contrast, in \cite{Brandino2011-thermo}, it was argued that the time for the sparse random ensemble scales like the volume, i.e., $t \sim \log d$.) . \\
{\bf Proof:}
From Theorem \ref{thm:main}  we need to compute the average: 
\be
\biggl\< \frac{|\eta|^2}{d^2}\frac{1}{d_S}+
\left( \frac{|\xi|^2}{d^2} - \frac{\gamma}{d^2} \right)^2 \biggr\>
\ee
over the distribution \eqref{eq:indep}. 
Straightforward calculations, relying on the assumption that the levels are independently, identically distributed
give:
\begin{eqnarray}
&& \<||\rho_S(t)-\omega_S||^{2}_{2}\>_H=\frac{\gamma}{d^2d_S}+\sum_{j\ne k}\frac{d_j^2d_k^2}{d^4}\\
&+& 2\sum_{j\ne k\ne s}\frac{d_j^2d_kd_s}{d^4}\left\<e^{iEt}\right\>^2
+\sum_{j\ne k\ne r\ne s}\frac{d_jd_kd_rd_s}{d^4}\left\<e^{iEt}\right\>^4\nonumber\\
&+& 2\sum_{j\ne k\ne s}\frac{d_j^2d_kd_s}{d^4}\left\<e^{2iEt}\right\> \left\<e^{iEt}\right\>^2
+ \sum_{j\ne k}\left(\frac{d_jd_k}{d^2d_S}+\frac{d_j^2d_k^2}{d^4}\right)\left\<e^{2iEt}\right\>^2.\nonumber
\end{eqnarray}
Substituting $\<e^{\pm iEt}\>=e^{-\sigma^2 t^2/2}$ we obtain:
\begin{eqnarray}
&& \<||\rho_S(t)-\omega_S||^{2}_{2}\>_H=\frac{\gamma}{d^2d_S}+\sum_{j\ne k}\frac{d_j^2d_k^2}{d^4}\label{av_gauss}\\
&+& 2\sum_{j\ne k\ne s}\frac{d_j^2d_kd_s}{d^4}e^{-t^2 \sigma^2}
+\sum_{j\ne k\ne r\ne s}\frac{d_jd_kd_rd_s}{d^4}e^{-2t^2 \sigma^2}\nonumber\\
&+& 2\sum_{j\ne k\ne s}\frac{d_j^2d_kd_s}{d^4}e^{-3t^2 \sigma^2}
+ \sum_{j\ne k}\left(\frac{d_jd_k}{d^2d_S}+\frac{d_j^2d_k^2}{d^4}\right)
e^{-4t^2 \sigma^2}.\nonumber
\end{eqnarray}

The assumed condition of no too big degeneracy (\ref{low_degen}) implies that: i) 
\be
\sum_{j\ne k}\frac{d_jd_k}{d^2}=O(1),\label{O(1)}
\ee 
which follows from the identity $1=\gamma/d^2+\sum_{j\ne k}d_jd_k/d^2$ and assumed $\gamma/d^2\ll 1$; ii) by the same reasoning:
\be
\sum_{j\ne k}\sum_{r\ne s}\frac{d_jd_kd_rd_s}{d^4}=O(1),\label{O(1)_2}
\ee
which follows from $1=\gamma^2/d^4+2(\gamma/d^2)\sum_{j\ne k}d_jd_k/d^2+\sum_{j\ne k}\sum_{r\ne s}d_jd_kd_rd_s/d^4$ and the first two term are $O(1/d^2)$ and $O(1/d)$ respectively; iii) $\sum_{j\ne k}d_j^2d_k^2/d^2<\gamma^2/d^4=O(1/d^2)$.

Thus the constant terms in (\ref{av_gauss}) are of the order $1/d$ and hence negligible. The lower bound in (\ref{thm_gauss}) is obtained by neglecting in (\ref{av_gauss}) everything but the leading part of the last term and using (\ref{O(1)}). To get the upper bound, we use (\ref{O(1)_2}) and substitute all the exponents in (\ref{av_gauss}) with the biggest one $e^{-t^2 \sigma^2}$.\blacksquare

\section{Conclusions}

We have shown that equilibration time of a small subsystem under the dynamics of a random Hamiltonian is fast, being determined by the mean inverse of the energy gaps of the Hamiltonian, which in typical cases scales as the number of particles in the system. This should be contrasted with the time scale that can be obtained from the results of \cite{Linden2009-thermo}, which is given by the inverse of the smallest energy gap of the Hamiltonian. The main message of this work is that in order to understand the equilibration time in quantum systems, one must consider more than the eigenvalues of the Hamiltonian. Indeed, the structure of the eigevectors of the Hamiltonian appears to be of crucial importance for equilibration to happen quickly. Interestingly, asymptotic equilibration can be inferred just from the knowledge of the eigenvalues of the model, this being the main result of \cite{Linden2009-thermo}.

In our work we have shown that for almost any choice of the eigenvectors (when picked from the Haar measure), the equilibration will happen quickly. A direct consequence of our result is that we can replace the Haar measure when choosing the basis by any quantum unitary 4-design, since we only used averages over four moments of the distribution in our arguments. As random quantum circuits of the order $n^{4}$ gates form a unitary 4-design \cite{Brandao-Harrow-Horodecki}, this means in particular that most Hamiltonians whose eigenbasis are determined by a sufficiently large quantum circuit (with more than $O(n^{4})$ gates) are such that small subsystems equilibrate fast. A drawback of the result is that typically a Hamiltonian chosen in this way will be very different from realistic Hamiltonians, which should be formed by a sum of few-body terms.

Comparing our result with other works, we want to say that a similar bound to this from Eq. \eqref{eq:main} was also obtained in \cite{Vinayak2011-thermo, Masanes2011-thermo, Cramer2011-thermo}, where in \cite{Cramer2011-thermo}, the author used his result to prove thermalization of some classes of local Hamiltonians. What is more, the time scale of the phenomena, obtained in these works, is similar to ours, namely, that the time is given by the Fourier transform of the function of the energy and that this time is, in fact, quite short.

In particular, using our approach, one can check that with high probability, the stationary state
of the system $\omega_S$ is close to the maximally mixed one. In a future work we aim to add
some locality constraints to the Hamiltonian in order to become closer to the thermodynamical regime,
where system is weakly coupled to the bath, so that it is meaningful to talk about a self Hamiltonian
of the system, and the latter would equilibrate to a Gibbs state determined by that Hamiltonian.

It is an interesting open problem if one can say something about the generic case of some more realistic type of models.

{\it Note added}: After the completion of this paper we became aware that similar results have been reported in \cite{Vinayak2011-thermo} and \cite{Masanes2011-thermo}.

{\it Acknowledgements. --}
We would like to thank Robert Alicki for stimulating discussions.
P.\'C., M.H., P.H., and J.K. are supported  by Polish Ministry of Science and Higher
Education grant N N202 231937. F.B. is supported by a "Conhecimento Novo" fellowship from the Brazilian agency Fundac\~ao de Amparo a Pesquisa do Estado de Minas Gerais (FAPEMIG). J.K. acknowledges the financial support of the QCS and TOQUATA projects. Part of this work was done at National Quantum Information Centre of Gda\'nsk. F.B. and M.H. acknowledge the hospitality of Mittag Leffler Institute within the program "Quantum Information Science", where part of the work was done. F.B. and J.K. acknowledge the kind hospitality of National Quantum Information Centre in Gdansk, where part of the work was completed.

\section{Appendix A: Proof of Proposition \ref{prop:dist-U4}}

We rewrite $\tr(\rho_S(t)-\omega_S)^2$ as follows (we will not put the dependence on time explicitly to shorten the notation)
\ben
&& \tr(\rho_{S} - \omega_{S})^{2} = \tr\rho_{S}^{2} - 2\tr\rho_{S}\omega_{S} + \tr\omega_{S}^{2} =
\nonumber \\
&&  \tr[(\rho_{S_1} \otimes \rho_{S_2} - \rho_{S_1} \otimes \omega_{S_2} - \omega_{S_1} \otimes \rho_{S_2} + \omega_{S_1} \otimes \omega_{S_1}) \mathbb{V}_{S_1:S_2}\bigr] \nonumber \\
&&= \tr(\rho_{1} \otimes \rho_{2}\, \mathbb{F}) - \tr(\rho_{1} \otimes \omega_{2}\, \mathbb{F}) -
\nonumber \\
 &&- \tr(\omega_{1} \otimes \rho_{2}\, \mathbb{F}) +\tr(\omega_{1} \otimes \omega_{2}\, \mathbb{F}),
 \label{eq:rho_s-omega_s}
\een
where the label $i=1,2,3,4$ denotes copies of the original system $S_iB_i$, so that e.g. $\rho_1=\rho_{S_1B_1}$.

Consider now the first term. Writing $\rho_{SB}=e^{-iHt} \sigma_{SB} e^{iHt}$
we obtain
\ben
&&\tr(\rho_{1} \otimes \rho_{2} \mbF) =  \tr(e^{-iHt} \sigma e^{iHt} \otimes e^{-iHt} \sigma e^{iHt}
\mbF) \nonumber \\
&&= \tr(\sigma_{1} \otimes \sigma_{2} W_{1} \otimes W_{2} \mbF W_{1}^\dagger \otimes W^\dagger_{2}).
\een
We can now use Eq.~(\ref{fac:ABCD}), putting $C_{12}=\sigma_{1} \otimes \sigma_{2}$,
$D_{12}=\mbF$, $A=B=W$. As a result we obtain
\ben
&&\tr(\rho_{1} \otimes \rho_{2} \mbF) =
\tr\bigl[\mbV_{12:34}
(W_{1} \otimes W_{2} \otimes W_{1}^\dagger \otimes W^\dagger_{2})\bigr]= \nonumber \\
&&=\tr\biggl[\mbV_{12:34}
U^{\ot 4} (W_0 \otimes W_0 \otimes W_0^\dagger \otimes W_0^\dagger) {U^\dagger}^{\ot 4}
\biggr].
\label{eq:YWWWW}
\een
In a similar way we get
\ben
&&\tr(\rho_{1} \otimes \omega_2 \mbF)=
\nonumber\\
&&=\sum_i \tr\biggl[\mbV_{12:34}
U^{\ot 4} (W_0 \otimes P_i \otimes W_0^\dagger \otimes P_i)
{U^\dagger}^{\ot 4}\biggr], \nonumber \\
&&\tr(\omega_1 \otimes \rho_2 \mbF)=
\nonumber\\
&&=\sum_i \tr\biggl[\mbV_{12:34}
U^{\ot 4} (P_i \otimes W_0 \otimes P_i \otimes W^\dagger_0)
{U^\dagger}^{\ot 4}\biggr] \nonumber \\
&& \tr(\omega_1 \otimes \omega_2 \mbF)=
\nonumber\\
&&=\sum_{ij} \tr\biggl[\mbV_{12:34}
U^{\ot 4} (P_i \otimes P_j \otimes P_i \otimes P_j)
{U^\dagger}^{\ot 4}\biggr].
\label{eq:YWP}
\een
If we now insert \eqref{eq:YWP} and \eqref{eq:YWWWW} into \eqref{eq:rho_s-omega_s}
we obtain the desired result \eqref{eq:dist}. $\blacksquare$

\section{Appendix b: Averages}
We prove here a few auxiliary facts.

\begin{proposition}
\label{fact:cycle}
For $\pi \in S_n$ being a cycle, we have
\be
\tr(V_\pi A_1\ot \ldots\ot A_n)=\tr (A_{\pi(1)}\ldots A_{\pi(n)})
\ee
\end{proposition}
{\bf Proof.} By direct inspection.

\begin{proposition}
Consider the twirling operation $\tau_n$ given by $\tau_n(\cdot) = \int d U U^{\ot n}(\cdot) {U^{\ot n}}^\dagger$.
Then for any operators $A$ and $B$ acting on $(C^d)^{\ot n}$  we have
\be
\tr [A \tau_n(B)]= \<\vec a |M^{-1} |\vec b\> \label{matrixM}
\ee
where $\vec a=(a_\pi)_{\pi\in S_n}$, $\vec b=(b_\pi)_{\pi\in S_n}$,
with $a_\pi=\tr A V_{\pi^{-1}}$,$b_\pi=\tr B V_{\pi^{-1}}$.
The matrix $M$ is given by $M_{\pi,\sigma}=\<V_\pi|V_\sigma\>=\tr (V_{\pi^{-1}} V_\sigma)$.
\label{fact:twirl}
\end{proposition}
{\bf Proof.}
It is easy to check that the twirling operation is an orthogonal
projector in the Hilbert-Schmidt space of operators, with the scalar product
$\<A|B\>=\tr (A^\dagger B)$. It projects onto
the space spanned by the permutation operators $V_\pi$. Then from Proposition \ref{lem:proj}
we have that
\be
\tr[A^\dagger \tau (B)]=\sum_{\pi,\sigma}\<A|V_\pi\> (M^{-1})_{\pi,\sigma}\<V_\sigma|B\>
\ee
However $\<A|V_\pi\>=\tr(A^\dagger V_\pi)=a_\pi^*$ and similarly $\<V_\sigma|B\>=b_\sigma$,
where $*$ stands for complex conjugate. This ends the proof.

\begin{proposition}
Let $\{\psi_i\}$  be an arbitrary set of vectors from the Hilbert space $\hcal$.
Let $M$ be the matrix of the elements from the set: $M_{ij}=\<\psi_i|\psi_j\>$, and let us denote by $M^{-1}$ the pseudoinverse of $M$, i.e. the unique matrix satisfying $M^{-1}M=MM^{-1}=Q$, where $Q$ is an orthogonal projection
onto a support of the matrix $M$ ($Q$ is the orthogonal projection onto the range of $M$). Then the orthogonal projector $P$ onto the subspace spanned by $\{\psi_i\}$ can be written as:
\be
P=\sum_{ij} X_{ij} |\psi_i\>\<\psi_j|,
\ee
where $X_{ij}$ are elements of matrix $X$ and by $X$ we mean $X = M^{-1}$, so the pseudoinverse of matrix $M$.
\label{lem:proj}
\end{proposition}
\noindent {\bf Proof.} \\
We must show that the operator $P$ is indeed an orthogonal projection i.e. that $P = P^{2}$.
Let start our proof by writing the following expression for the $P^{2}$:
\ben
P^{2} &&= \nonumber \sum_{ijkl} X_{ij} X_{kl} |\psi_i\>\<\psi_j|\psi_k\>\<\psi_l| \nonumber \\
&&= \sum_{ijkl} X_{ij} X_{kl} |\psi_i\>\<\psi_l|M_{jk}
\nonumber \\
&&= \sum_{ijl} X_{ij} |\psi_i\>\<\psi_l| \sum_{k} M_{jk}X_{kl},
\een
where we use definition of $M$ from Proposition \ref{lem:proj}.
We can now express our equation in terms of $Q$ and use this to obtain the desired result
\ben
P^{2} &&= \sum_{ijl} X_{ij} |\psi_i\>\<\psi_l| Q_{jl}
= \sum_{il} \left(\sum_{j} X_{ij} Q_{jl}\right) |\psi_i\>\<\psi_l| \nonumber \\
&&= \sum_{il} (XQ)_{il} |\psi_i\>\<\psi_l| = \sum_{il} X_{il} |\psi_i\>\<\psi_l| = P,
\een
since according to Proposition \ref{lem:proj} $MQ=QM=M$ and $XQ=QM=X$. $\blacksquare$

\section{Appendix C: Inverse of the matrix $M$}
\label{Ap:B}
In this section we derive properties of the matrix $M$ which were needed in the proof of Theorem \eqref{thm:main}.

\subsection{Properties of $M$ matrix for general representations}
We will first introduce some notation.
Denote by $G$ an arbitrary finite group, $\left\vert G\right\vert =n$. Let
\be
D^{\alpha }:G\rightarrow Hom(\hcal^{\alpha }); \ \alpha =1,2,....,r;
\dim \hcal^{\alpha } = d_{\alpha }
\ee%
be all inequivalent, irreducible representations $(IR)$ (not necessarily
unitary) of $G$ and let
\be
D^{\alpha }(g) = (D_{ij}^{\alpha }(g)); \ i,j=1,2,....,d_{\alpha }
\ee%
be their matrix forms where $D^{1}(g)=1$ is the trivial representation. By
\be
\chi ^{\alpha }(g) = \tr(D_{ij}^{\alpha }(g))
\ee%
we denote the corresponding irreducible character $(ICH).$
We now define our main object - the matrix $M^D$.
\begin{definition}
Let $D:G$ $\rightarrow Hom(\hcal)$ be any representation (not necessarily
unitary) of $\ G$. Define a matrix $M\in M(n,%
\mathbb{C}
)$
\ben
M^{D}&=&(m_{gh}) = (\tr(D^{-1}(g)D(h))) \nonumber \\
&=& (\tr(D(g^{-1}h))) = (\chi ^{D}(g^{-1}h))
\een
\end{definition}
We apply this definition to irreducible representations $D^\alpha$:
\begin{definition}
For irreducible representations $D^{\alpha }$ we define the corresponding matrices%
\ben
M^{\alpha } &=& (m_{gh}^{\alpha }) = (\tr(D^{\alpha })^{-1}(g)D^{\alpha
}(h))) \nonumber \\
&=& (\tr(D^{\alpha }(g^{-1}h)))=(\chi ^{\alpha }(g^{-1}h))
\een
\end{definition}

Thus from the definition of $M^{\alpha }$, it follows that in order to
calculate the entries of $M^{\alpha }$ we do not need to know explicitly irrep
\ $D^{\alpha }$, but only $ICH$ \ $\chi ^{\alpha }$.

Now we shall express the matrix $M^D$ by means of the matrices $M^\alpha$.
Namely, from the decompositions
\be
D=\oplus _{\alpha =1}^{r}k_{\alpha }D^{\alpha }; \ k_{\alpha }\in
\mathbb{N}
\cup \{0\} \quad \Rightarrow \quad \chi ^{D}=\sum_{\alpha =1}^{r}k_{\alpha }\chi
^{\alpha },
\ee%
where $k_{\alpha }$ is the multiplicity of irrep \ $D^{\alpha }$ in $D$ and
from the character properties we get

\begin{proposition}
1.Matrices $M^{\alpha }$ are Hermitian and%
\be
M^{D}=\sum_{\alpha =1}^{r}k_{\alpha }M^{\alpha }\quad \Rightarrow \quad
(M^{D})^{+}=M^{D}.
\ee

2.The sum of elements in each row and column of the matrix $M^{D}$ is equal
to $nk_{1}.$
\end{proposition}

Further, using orthogonality relations for $ICH$
\be
\frac1n \sum_{g\in G}\chi^\alpha(g) \chi^\beta(g^{-1})=\delta_{\alpha\beta},
\ee
which one can derive from Schur's lemma,
one can prove
\begin{proposition}
\label{propP}
The matrices $M^{\alpha }$ are proportional to orthogonal projectors:
\be
M^{\alpha }M^{\beta }=\frac{n}{d_{\alpha }}\delta ^{\alpha \beta }M^{\alpha
}
\ee%
whereas the matrices $P^{\alpha }=\frac{d_{\alpha }}{n}M^{\alpha }$ form the
complete set of orthogonal projectors:
\be
P^{\alpha }P^{\beta }=\delta ^{\alpha \beta }P^{\alpha }; \ \sum_{\alpha
=1}^{r}P^{\alpha } = \mathbf{1}; \ \left( P^{\alpha }\right) ^{+}=P^{\alpha
}
\ee%
In particular the matrices $M^{\alpha }$ and $P^{\alpha }$ mutually commute.
\end{proposition}

This already gives us eigenvalues of the matrix $M^D$ in terms of
dimensions $d_\alpha$ and multiplicities of the irreps, which allows us to
derive the formula for the inverse of $M_D$, whenever it exists (see Theorem \ref{thm:poly}).
We can however also find eigenvectors in terms of matrix elements of irreps.
Namely, consider $n$ vectors in $\mathbb{C}^{n}$ whose entries are
defined by the matrix elements of irrep \ $D^{\alpha }$ in the following way:
\ben
&U_{ij}^{\alpha }&=(D_{ij}^{\alpha }(g^{-1})) \in
\mathbb{C}^{n}; g \in G; \alpha =1,2,....,r; \nonumber \\
&&i,j=1,2,....,d_{\alpha }.
\een%
where $\alpha ,\ i,$ $j$ label the vectors $U_{ij}^{\alpha }$ and $g\in G$
label the entries of the vector $U_{ij}^{\alpha }\in
\mathbb{C}^{n}$, i.e., the vector $U_{ij}^{\alpha }$ has the form%
\be
\left( U_{ij}^{\alpha }\right) ^{T}=(D_{ij}^{\alpha
}(g_{1}^{-1}), D_{ij}^{\alpha }(g_{2}^{-1}),...,D_{ij}^{\alpha
}(g_{n}^{-1}))\in
\mathbb{C}^{n},
\ee%
and in particular%
\be
\left( U^{1}\right) ^{T}=(1,1,...,1)\in
\mathbb{C}
^{n}
\ee%
It turns out that these vectors are eigenvectors of the matrices $M^\alpha$:
\begin{proposition}
The $U_{ij}^{\alpha }$ are linearly independent and they are eigenvectors
for matrices $M^{\alpha }$ and $P^{\alpha }$; i.e.,
\be
M^{\alpha }U_{ij}^{\beta }=\delta ^{\alpha \beta }\frac{n}{d_{\alpha }}%
U_{ij}^{\beta }; \ P^{\alpha }U_{ij}^{\beta }=\delta ^{\alpha \beta
}U_{ij}^{\beta }
\ee%
If the irrep $D^{\alpha }$ are unitary then the vectors $U_{ij}^{\alpha }$
are orthogonal with respect to the standard scalar product in $\mathbb{C}^{n}.$
\label{prop:eigenvectors}
\end{proposition}
{\bf Proof:}
In order to prove this Proposition we will need:

\begin{proposition}
Let $\chi :G\rightarrow
\mathbb{C}
$ be any character of the group $G$ (or even any central function on $G$)
and $D^{\alpha }$ be an irrep of $G$. Then%
\be
\Phi :\hcal^{\alpha } \rightarrow \hcal^{\alpha }; \ \Phi _{ij}=\sum_{g\in G}%
\overline{\chi (g)}D_{ij}^{\alpha }(g)=\frac{n}{d_{\alpha }}(\chi ^{\alpha
},\chi )\delta _{ij},
\ee%
where $(\cdot ,\cdot)$ is a scalar product in the space $%
\mathbb{C}
^{G}.$
\end{proposition}

Now we can prove Proposition \ref{prop:eigenvectors}.

{\bf Proof.}
\be
(M^{\alpha }U_{ij}^{\beta })_{g}=\sum_{h}\chi ^{\alpha
}(g^{-1}h)D_{ij}^{\beta }(h^{-1}).
\ee%
We set%
\be
u^{-1}=g^{-1}h
\ee%
then%
\be
(M^{\alpha }U_{ij}^{\beta })_{g}=\sum_{u}\overline{\chi ^{\alpha }(u)}%
D_{in}^{\beta }(u)D_{nj}^{\beta }(g^{-1}).
\ee%
Now we use the above proposition and the fact that $ICH$ of $G$ are orthonormal,
i.e., $(\chi ^{\alpha },\chi ^{\beta })=\delta ^{\alpha \beta }$, and we get
\ben
(M^{\alpha }U_{ij}^{\beta })_{g}&=&\sum_{n}\delta ^{\alpha \beta }\frac{n}{%
d_{\alpha }}\delta _{in}D_{nj}^{\beta }(g^{-1}) \nonumber \\
&=& \delta ^{\alpha \beta }\frac{%
n}{d_{\alpha }}D_{ij}^{\beta }(g^{-1}) \nonumber \\
&=& \delta ^{\alpha \beta }\frac{n}{%
d_{\alpha }}(U_{ij}^{\beta })_{g}
\een

As an easy corollary from Proposition \ref{prop:eigenvectors} we get the following
theorem concerning the eigenproblem for the matrix $M^{D}$:

\begin{theorem}
\label{thm:spectral}
The vectors \ $U_{ij}^{\beta }$ \ are eigenvectors for the matrix $M^{D}$, i.e.,
\be
M^{D}U_{ij}^{\beta }=k_{\beta }\frac{n}{d_{\beta }}U_{ij}^{\beta
},
\ee
and the eigenvalues of $M^{D}$ are the following:
\be
\lambda _{\beta }\equiv k_{\beta }\frac{n}{d_{\beta }}.
\ee
The spectral decomposition of $M^D$ thus reads
\be
M^D=\sum_{\alpha=1}^r  \lambda_\alpha P^\alpha
\ee
where the eigenprojectors $P^\alpha$ are defined in Proposition \ref{propP}.
\end{theorem}

Directly from this theorem follows

\begin{corollary}
\label{cor:inv}
1. The matrix $M^{D}$ is invertible iff each multiplicity $k_{\alpha
}$ in the decomposition
\be
\chi ^{D}=\sum_{\alpha =1}^{r}k_{\alpha }\chi ^{\alpha }\quad
\Leftrightarrow \quad M^{D}=\sum_{\alpha =1}^{r}k_{\alpha }M^{\alpha }
\ee%
is nonzero.

2. For a given $\alpha $ the vectors $U_{ij}^{\alpha }\qquad
i,j=1,2,....,d_{\alpha }$ span the eigenspace for the eigenvalue $\lambda
_{\alpha },$ so the multiplicity of $\lambda _{\alpha }$ is equal to $%
d_{\alpha }^{2}.$

3. The eigenvectors $U_{ij}^{\alpha }$ do not depend on the representation
$D:G$ $\rightarrow Hom(V),$ whereas the eigenvalues $\lambda _{\alpha }$
depend on the representation $D:G$ $\rightarrow Hom(V)$ via multiplicities $%
k_{\alpha }.$

4. We have also
\begin{eqnarray*}
\det M^{D} &=&\Pi _{_{\alpha =1}}^{r}(k_{\alpha }\frac{n}{d_{\alpha }}%
)^{d_{\alpha }^{2}}, \\
\tr M^{D} &=&\sum_{\alpha =1}^{r}nk_{\alpha }d_{\alpha }=n\dim D
\end{eqnarray*}
\end{corollary}

Thus in order to calculate the eigenvalues $\lambda _{\alpha }$ of
the matrix $M^{D}$ we need only the multiplicities $k_{\alpha }$ of irrep \ $%
D^{\alpha }$ in the representation $D$ (the dimensions $d_{\alpha }$ and
rank $n=|G|$ are known).

From the above spectral decomposition we get

\begin{corollary}
If the matrix \ $M^{D}=\sum_{\alpha =1}^{r}k_{\alpha }M^{\alpha }$ is
invertible ($\Leftrightarrow \quad k_{\alpha }\geq 1)$ then%
\be
(M^{D})^{-1}=\sum_{\alpha =1}^{r}\lambda _{\alpha }^{-1}P^{\alpha
}=\sum_{\alpha =1}^{r}\frac{d_{\alpha }}{nk_{\alpha }}P^{\alpha }=\frac{1}{%
n^{2}}\sum_{\alpha =1}^{r}\frac{d_{\alpha }^{2}}{k_{\alpha }}M^{\alpha
}
\ee%
In fact this formula expresses the entries of the matrix $(M^{D})^{-1}$ in
terms of $ICH$; namely we have
\be
(M^{D})_{gh}^{-1}=\frac{1}{n^{2}}\sum_{\alpha =1}^{r}\frac{d_{\alpha }^{2}}{%
k_{\alpha }}\chi ^{\alpha }(g^{-1}h),
\ee%
i.e., all we need to calculate $(M^{D})^{-1}$ are $ICH$ and the
multiplicities $k_{\alpha }$ of irrep $D^{\alpha }$ in the representation $D$.
\end{corollary}

\begin{remark}
\label{remarkprop}
It is known \cite{Fulton1991-book-rep} that one can calculate the multiplicities $k_{\alpha }$
of irrep $D^{\alpha }$ in an arbitrary representation $R$ of the group $G$
using the following formula:
\be
k_{\alpha }=(\chi ^{R},\chi ^{\alpha })\equiv \frac{1}{n}\sum_{g\in G}\chi
^{R}(g)\chi ^{\alpha }(g^{-1}),
\ee%
where $(\chi ^{R},\chi ^{\alpha })$ is the scalar product in the linear
space of central functions on the group $G.$
\end{remark}

Finally, we want to express the inverse of $M^D$ as a polynomial of $M^D$.
To this end, note that from the Hermiticity of the matrix $M^{D}$ it follows
that the rank of the minimal polynomial of $M^{D}$ is equal to $r$ and the coefficients
of this polynomial are determined by $r$ pairwise distinct eigenvalues of $M^{D}$.
Thus it is possible to write the matrix $ (M^{D})^{-1}$ as a polynomial of degree $r-1$ in $M^{D}.$ In fact we have

\begin{theorem}
\label{thm:poly}
Let
\be
W(x)=x^{r}+s_{r-1}x^{r-1}+...+s_{1}x+s_{0}
\ee%
be a minimal polynomial of the matrix $M^{D}$; i.e., $W(M^{D})=0.$ Then if $%
s_{0}\neq 0$,
\be
(M^{D})^{-1}=\frac{-1}{s_{0}}%
[(M^{D})^{r-1}+s_{r-1}(M^{D})^{r-2}+...+s_{2}M^{D}+s_{1}]
\ee%
This formula expresses the inverse of the matrix $M^{D}$ as a polynomial
function of itself.
\end{theorem}
In the next section we shall apply these results to our representation.

\subsection{Applications}

In this subsection we will apply the above results to a particular
representation of the symmetric group $S_{n}$ 

\begin{definition}
Let $H=\otimes _{i=1}^{n}%
\mathbb{C}
^{d},$ so $\dim H=d^{n}.$ We define the representation $D$ of the group $%
S_{n}$ in the space $H$ by means of operators which swap subsystems:
\ben
\forall \sigma \in S_{n}\quad &D&(\sigma )(e_{i_{1}}\otimes e_{i_{2}}\otimes
...\otimes e_{i_{n}}) \nonumber \\
&=&e_{\sigma ^{-1}(i_{1})}\otimes e_{\sigma
^{-1}(i_{2})}\otimes ...\otimes e_{\sigma ^{-1}(i_{n})}
\een%
where $\{e_{i}\}_{i=1}^{d}$ is a basis of $%
\mathbb{C}^{d}.$
\end{definition}
In other words $D(\sigma)=V_\sigma$, using notation from previous sections.

An important property of any representation is its character and
in this case it is not very difficult to prove that

\begin{proposition}
The character of the representation $D:S_{n}\rightarrow Hom(H)$  has the
following form:
\be
\forall \sigma \in S_{n}\quad \chi^{D}(\sigma )=d^{l(\sigma )},
\ee%
where $l(\sigma )$ is the number of cycles in the cycle decomposition of $%
\sigma \in S_{n}.$
\end{proposition}

It follows that in the case of the representation $D$ of $S_{n}$ the matrix $%
M^{D}$ has the form
\be
M^{D}=(m_{\sigma \pi })=(\chi ^{D}(\sigma ^{-1}\pi ))=(d^{l(\sigma ^{-1}\pi
)})
\ee

\begin{example}
For the group $S_{3}$ the matrix $M^{D}$ is the following:
\be
M^{D}=\left(
\begin{array}{cccccc}
d^{3} & d^{2} & d^{2} & d^{2} & d & d \\
d^{2} & d^{3} & d & d & d^{2} & d^{2} \\
d^{2} & d & d^{3} & d & d^{2} & d^{2} \\
d^{2} & d & d & d^{3} & d^{2} & d^{2} \\
d & d^{2} & d^{2} & d^{2} & d^{3} & d \\
d & d^{2} & d^{2} & d^{2} & d & d^{3}%
\end{array}%
\right)
\ee
\end{example}

From Theorem \ref{thm:spectral} and Corollary \ref{cor:inv} of the previous subsection it follows that in
order to describe the basic properties of the matrix $M^{D}$, in particular
its eigenvalues and the inverse $(M^{D})^{-1}$, one has to calculate the
multiplicities $k_{\alpha }$ of irrep $D^{\alpha }$ in the representation $D$. Using
the formula from Remark \ref{remarkprop} and the character tables for $S_{3}$
and $S_{4}$ \cite{Fulton1991-book-rep} one gets

\begin{proposition}
1. The multiplicity coefficients $k_{\alpha }$ for $S_{3}$ are the following:
\be
k_{1}=\frac{1}{6}(d^{3}+3d^{2}+2d); \ k_{2}=\frac{1}{6}%
(d^{3}-3d^{2}+2d);  k_{3}=\frac{1}{3}(d^{3}-d)
\ee

2. The multiplicity coefficients $k_{\alpha }$ \ in the case of $S_{4}$ are
of the form%
\ben
k_{1} &=&\frac{1}{4!}d(d+1)(d+2)(d+3); \nonumber \\
k_{2} &=& \frac{1}{4!}%
d(d-1)(d-2)(d-3); \nonumber \\
k_{3} &=& \frac{2}{4!}d^{2}(d^{2}-1); \nonumber \\
k_{4} &=&\frac{3}{4!}d(d^{2}-1)(d-2); \nonumber \\
k_{5} &=&\frac{3}{4!}%
d(d^{2}-1)(d+2)
\een
\end{proposition}

From Theorem \ref{thm:spectral} we get immediately the values of the corresponding eigenvalues
and then from Corollary \ref{cor:inv} and Theorem \ref{thm:poly} we get

\begin{theorem}
For $S_{3}$ we have%
\be
M^{-1}=\frac{1}{d^{3}(d^{2}-1)^{2}(d^{2}-4)}%
(M^{2}-3d(d^{2}+1)M+3d^{4}(d^{2}-1)\mathbf{1})
\ee

where $d \neq 1,2$ and
\ben
&&M^{-1} = \frac{1}{s_{3}} \times \nonumber \\ \label{eq:Matrix}
&&\left(
\begin{array}{cccccc}
a_{11} & a_{12} & a_{13} & a_{14} &
a_{15}& a_{16} \\
a_{21} & a_{22} &  a_{23}  & a_{24}
& a_{25} & a_{26} \\
a_{31} & a_{32} & a_{33} & a_{34} & a_{35} &
a_{36}\\
a_{41} & a_{42} & a_{43} & a_{44} & a_{45}
& a_{46}\\
a_{51} & a_{52} & a_{53} & a_{54} &
a_{55}& a_{56} \\
a_{61} & a_{62} & a_{63} & a_{64} & a_{65} & a_{66}
\end{array}
\right),
\een
where \\
$a_{11} = d^{6}-3d^{4}+2d^{2}; \ a_{12} = d^{3}-d^{5}; \ a_{13} = d^{3}-d^{5}; \\
a_{14} = d^{3}-d^{5}; \ a_{15} = 2d^{4}-2d^{2}; \ a_{16} = 2d^{4}-2d^{2}; \\
a_{21} = d^{3}-d^{5}; \ a_{22} = d^{6}-3d^{4}+2d^{2}; \ a_{23} = 2d^{4}-2d^{2}; \\
a_{24} = 2d^{4}-2d^{2}; \ a_{25} = d^{3}-d^{5}; \ a_{26} = d^{3}-d^{5}; \\
a_{31} = d^{3}-d^{5}; \ a_{32} = 2d^{4}-2d^{2}; \ a_{33} = d^{6}-3d^{4}+2d^{2}; \\
a_{34} = 2d^{4}-2d^{2}; \ a_{35} = d^{3}-d^{5}; \ a_{36} = d^{3}-d^{5}; \\
a_{41} = d^{3}-d^{5}; \ a_{42} = 2d^{4}-2d^{2}; \ a_{43} = 2d^{4}-2d^{2}; \\
a_{44} = d^{6}-3d^{4}+2d^{2}; \ a_{45} = d^{3}-d^{5}; \ a_{46} = d^{3}-d^{5}; \\
a_{51} = 2d^{4}-2d^{2}; \ a_{52} = d^{3}-d^{5}; \ a_{53} = d^{3}-d^{5}; \\
a_{54} = d^{3}-d^{5}; \ a_{55} = d^{6}-3d^{4}+2d^{2}; \ a_{56} =  2d^{4}-2d^{2}; \\
a_{61} = 2d^{4}-2d^{2}; \ a_{62} = d^{3}-d^{5}; \ a_{63} = d^{3}-d^{5}; \\
a_{64} =  d^{3}-d^{5}; \ a_{65} = 2d^{4}-2d^{2}; \ a_{66} = d^{6}-3d^{4}+2d^{2}$
\end{theorem}
and $s_{3}=d^{3}(d^{2}-1)^{2}(d^{2}-4)=\allowbreak
9d^{5}-4d^{3}-6d^{7}+d^{9}$

In a similar way we obtain the result we used to prove Theorem~\ref{thm:main}

\begin{theorem}
For $S_{4}$ we have
\be
M^{-1}=\frac{1}{s_{5}}(M^{4}-s_{1}M^{3}+s_{2}M^{2}-s_{3}M^{1}+s_{4}\mathbf{1)},
\label{eq:inverse-M}
\ee

where $d \neq 1,2,3$ and
\ben
s_{1} &=& d^{2}(5d^{2}+19); \nonumber \\
s_{2} &=& 2d^{2}(d^{2}-1)(5d^{4}+23d^{2}+20); \nonumber \\
s_{3} &=& 2d^{4}(d^{2}-1)^{2}(5d^{4}+7d^{2}+12); \nonumber \\
s_{4} &=& d^{4}(d^{2}-1)^{3}(d^{2}-4)(5d^{4}-9d^{2}+36); \nonumber \\
s_{5} &=& d^{6}(d^{2}-1)^{4}(d^{2}-4)^{2}(d^{2}-9).
\een

\end{theorem}

\subsection{Miscellaneous facts about matrix $M^D$}
It turns out that the matrix $M^{D}$ may be written as a linear combination of adjacency matrices of the
so called Commutative Association Scheme (see \cite{Bannai1984-CAS}) determined
by the class structure of the group $G.$

\begin{definition}
Let $C_{1}=\{e\},$ $C_{2,}....,C_{r}$ be the conjugacy classes of the group $%
G.$ We define the $i^{th}$ relation $R_{i}$ on $G\times G$ in the following
way:
\[
(g,h)\in R_{i}\quad \Leftrightarrow \quad g^{-1}h\in C_{i}.
\]%
Then the pair $(G,\{R_{i}\}_{i=1}^{r})$ is a Commutative Assotiation Scheme
and by $A_{i}$ we denote the corresponding adjacency matrices which are
matrices of degree $|G|=n$ whose rows and columns are indexed by the
elements $G$ and whose entries are
\[
(A_{i})_{(g,h)}=%
\begin{array}{c}
1\quad if\quad (g,h)\in R_{i} \\
0\quad if\quad (g,h)\notin R_{i}%
\end{array}%
.
\]%
So $i$'th adjacency matrix $A_{i}$ is a $0,1$ matrix.
\end{definition}

\begin{proposition} \cite{Bannai1984-CAS}
\qquad (i) $A_{1}=\mathbf{1},$ the identity matrix.

(ii) $\sum_{k=1}^{r} A_{k}=J,$ where $J$ is the matrix whose entries are all
$1.$

(iii) $A_{k}^{t}=A_{k^{\prime }}$ for some $k^{\prime }\in \{1,...,r\}.$

(iv) $A_{i}A_{j}=\sum_{k=1}^{r}p_{ij}^{k}A_{k}$ $\ \forall i,j,k\in
\{1,...,r\}.$

(v) $p_{ij}^{k}=p_{ji}^{k}\quad \forall i,j,k\in \{1,...,r\}$ \ $%
\Leftrightarrow $ \ $A_{i}A_{j} =$ \\
$A_{j}A_{i}\quad \forall i,j\in
\{1,...,r\}.$
\end{proposition}

The matrix $M^{D}$ may be written as a linear combination of the adjacency
matrices in the following way

\begin{proposition}
\[
M^{D}=\sum_{i=1}^{r}\chi ^{D}(C_{i})A_{i}.
\]
\end{proposition} \blacksquare

\bibliographystyle{apsrev}

\bibliography{rmp15-hugekey}

\begin{thebibliography}{30}
\expandafter\ifx\csname natexlab\endcsname\relax\def\natexlab#1{#1}\fi
\expandafter\ifx\csname bibnamefont\endcsname\relax
  \def\bibnamefont#1{#1}\fi
\expandafter\ifx\csname bibfnamefont\endcsname\relax
  \def\bibfnamefont#1{#1}\fi
\expandafter\ifx\csname citenamefont\endcsname\relax
  \def\citenamefont#1{#1}\fi
\expandafter\ifx\csname url\endcsname\relax
  \def\url#1{\texttt{#1}}\fi
\expandafter\ifx\csname urlprefix\endcsname\relax\def\urlprefix{URL }\fi
\providecommand{\bibinfo}[2]{#2}
\providecommand{\eprint}[2][]{\url{#2}}

\bibitem[{\citenamefont{{Goldstein}
  et~al.}(2010{\natexlab{a}})\citenamefont{{Goldstein}, {Lebowitz}, {Tumulka},
  and {Zangh{\`i}}}}]{Goldstein2010-vonNeumann}
\bibinfo{author}{\bibfnamefont{S.}~\bibnamefont{{Goldstein}}},
  \bibinfo{author}{\bibfnamefont{J.~L.} \bibnamefont{{Lebowitz}}},
  \bibinfo{author}{\bibfnamefont{R.}~\bibnamefont{{Tumulka}}},
  \bibnamefont{and}
  \bibinfo{author}{\bibfnamefont{N.}~\bibnamefont{{Zangh{\`i}}}},
  \bibinfo{journal}{Eur. Phys. J. H} \textbf{\bibinfo{volume}{35}},
  \bibinfo{pages}{173} (\bibinfo{year}{2010}{\natexlab{a}}),
  \eprint{arXiv:1003.2129}.

\bibitem[{\citenamefont{Gemmer et~al.}(2004)\citenamefont{Gemmer, Michel, and
  Mahler}}]{Mahler2004-thermo}
\bibinfo{author}{\bibfnamefont{J.}~\bibnamefont{Gemmer}},
  \bibinfo{author}{\bibfnamefont{M.}~\bibnamefont{Michel}}, \bibnamefont{and}
  \bibinfo{author}{\bibfnamefont{G.}~\bibnamefont{Mahler}},
  \emph{\bibinfo{title}{Quantum Thermodynamics}}, vol. \bibinfo{volume}{657} of
  \emph{\bibinfo{series}{Lecture Notes in Physics}}
  (\bibinfo{publisher}{Springer}, \bibinfo{address}{Berlin},
  \bibinfo{year}{2004}).

\bibitem[{\citenamefont{{Tasaki}}(1998)}]{Tasaki1998-thermo}
\bibinfo{author}{\bibfnamefont{H.}~\bibnamefont{{Tasaki}}},
  \bibinfo{journal}{Phys. Rev. Lett.} \textbf{\bibinfo{volume}{80}},
  \bibinfo{pages}{1373} (\bibinfo{year}{1998}),
  \eprint{arXiv:cond-mat/9707253}.

\bibitem[{\citenamefont{{Linden} et~al.}(2009)\citenamefont{{Linden},
  {Popescu}, {Short}, and {Winter}}}]{Linden2009-thermo}
\bibinfo{author}{\bibfnamefont{N.}~\bibnamefont{{Linden}}},
  \bibinfo{author}{\bibfnamefont{S.}~\bibnamefont{{Popescu}}},
  \bibinfo{author}{\bibfnamefont{A.~J.} \bibnamefont{{Short}}},
  \bibnamefont{and} \bibinfo{author}{\bibfnamefont{A.}~\bibnamefont{{Winter}}},
  \bibinfo{journal}{Phys. Rev. E} \textbf{\bibinfo{volume}{79}},
  \bibinfo{pages}{061103} (\bibinfo{year}{2009}), \eprint{arXiv:0812.2385}.

\bibitem[{\citenamefont{{Goldstein}
  et~al.}(2010{\natexlab{b}})\citenamefont{{Goldstein}, {Lebowitz},
  {Mastrodonato}, {Tumulka}, and {Zanghi}}}]{Goldstein2010-thermo}
\bibinfo{author}{\bibfnamefont{S.}~\bibnamefont{{Goldstein}}},
  \bibinfo{author}{\bibfnamefont{J.~L.} \bibnamefont{{Lebowitz}}},
  \bibinfo{author}{\bibfnamefont{C.}~\bibnamefont{{Mastrodonato}}},
  \bibinfo{author}{\bibfnamefont{R.}~\bibnamefont{{Tumulka}}},
  \bibnamefont{and} \bibinfo{author}{\bibfnamefont{N.}~\bibnamefont{{Zanghi}}},
  \bibinfo{journal}{\pre} \textbf{\bibinfo{volume}{81}},
  \bibinfo{pages}{011109} (\bibinfo{year}{2010}{\natexlab{b}}),
  \eprint{arXiv:0911.1724}.

\bibitem[{\citenamefont{Reimann}(2008)}]{Reimann2008-thermo}
\bibinfo{author}{\bibfnamefont{P.}~\bibnamefont{Reimann}},
  \bibinfo{journal}{Phys. Rev. Lett.} \textbf{\bibinfo{volume}{101}},
  \bibinfo{pages}{190403} (\bibinfo{year}{2008}), \eprint{arXiv:0810.3092}.

\bibitem[{\citenamefont{{Gong} and {Duan}}(2011)}]{Gong2011}
\bibinfo{author}{\bibfnamefont{Z.-X.} \bibnamefont{{Gong}}} \bibnamefont{and}
  \bibinfo{author}{\bibfnamefont{L.}~\bibnamefont{{Duan}}},
  \emph{\bibinfo{title}{{Comment on ''Foundation of Statistical Mechanics under
  Experimentally Realistic Conditions''}}} (\bibinfo{year}{2011}),
  \eprint{arXiv:1109.4696}.

\bibitem[{\citenamefont{Rigol et~al.}(2008)\citenamefont{Rigol, Dunjko, and
  Olshanii}}]{Rigol2008-thermo}
\bibinfo{author}{\bibfnamefont{M.}~\bibnamefont{Rigol}},
  \bibinfo{author}{\bibfnamefont{V.}~\bibnamefont{Dunjko}}, \bibnamefont{and}
  \bibinfo{author}{\bibfnamefont{M.}~\bibnamefont{Olshanii}},
  \bibinfo{journal}{Nature} \textbf{\bibinfo{volume}{452}},
  \bibinfo{pages}{854} (\bibinfo{year}{2008}), \eprint{arXiv:0708.1324}.

\bibitem[{\citenamefont{Rigol}(2009)}]{Rigol2009-thermo}
\bibinfo{author}{\bibfnamefont{M.}~\bibnamefont{Rigol}},
  \bibinfo{journal}{Phys. Rev. Lett.} \textbf{\bibinfo{volume}{103}},
  \bibinfo{pages}{100403} (\bibinfo{year}{2009}), \eprint{arXiv:0904.3746}.

\bibitem[{\citenamefont{Cassidy et~al.}(2011)\citenamefont{Cassidy, Clark, and
  Rigol}}]{Cassidy2011-thermo}
\bibinfo{author}{\bibfnamefont{A.~C.} \bibnamefont{Cassidy}},
  \bibinfo{author}{\bibfnamefont{C.~W.} \bibnamefont{Clark}}, \bibnamefont{and}
  \bibinfo{author}{\bibfnamefont{M.}~\bibnamefont{Rigol}},
  \bibinfo{journal}{Phys. Rev. Lett.} \textbf{\bibinfo{volume}{106}},
  \bibinfo{pages}{140405} (\bibinfo{year}{2011}), \eprint{arXiv:1008.4794}.

\bibitem[{\citenamefont{Banuls et~al.}(2011)\citenamefont{Banuls, Cirac, and
  Hastings}}]{Banuls2011-thermo}
\bibinfo{author}{\bibfnamefont{M.~C.} \bibnamefont{Banuls}},
  \bibinfo{author}{\bibfnamefont{J.~I.} \bibnamefont{Cirac}}, \bibnamefont{and}
  \bibinfo{author}{\bibfnamefont{M.~B.} \bibnamefont{Hastings}},
  \bibinfo{journal}{Phys. Rev. Lett.} \textbf{\bibinfo{volume}{106}},
  \bibinfo{pages}{050405} (\bibinfo{year}{2011}), \eprint{arXiv:1007.3957}.

\bibitem[{\citenamefont{Cramer and Eisert}(2010)}]{Cramer2010-thermo}
\bibinfo{author}{\bibfnamefont{M.}~\bibnamefont{Cramer}} \bibnamefont{and}
  \bibinfo{author}{\bibfnamefont{J.}~\bibnamefont{Eisert}},
  \bibinfo{journal}{New J. Phys.} \textbf{\bibinfo{volume}{12}},
  \bibinfo{pages}{055020} (\bibinfo{year}{2010}), \eprint{arXiv:0911.2475}.

\bibitem[{\citenamefont{Gogolin et~al.}(2011)\citenamefont{Gogolin, Mueller,
  and Eisert}}]{Gogolin2011-thermo}
\bibinfo{author}{\bibfnamefont{C.}~\bibnamefont{Gogolin}},
  \bibinfo{author}{\bibfnamefont{M.~P.} \bibnamefont{Mueller}},
  \bibnamefont{and} \bibinfo{author}{\bibfnamefont{J.}~\bibnamefont{Eisert}},
  \bibinfo{journal}{Phys. Rev. Lett.} \textbf{\bibinfo{volume}{106}},
  \bibinfo{pages}{040401} (\bibinfo{year}{2011}), \eprint{arXiv:1009.2493}.

\bibitem[{\citenamefont{Cramer et~al.}(2008)\citenamefont{Cramer, Dawson,
  Eisert, and Osborne}}]{Cramer2008-thermo}
\bibinfo{author}{\bibfnamefont{M.}~\bibnamefont{Cramer}},
  \bibinfo{author}{\bibfnamefont{C.~M.} \bibnamefont{Dawson}},
  \bibinfo{author}{\bibfnamefont{J.}~\bibnamefont{Eisert}}, \bibnamefont{and}
  \bibinfo{author}{\bibfnamefont{T.~J.} \bibnamefont{Osborne}},
  \bibinfo{journal}{Phys. Rev. Lett.} \textbf{\bibinfo{volume}{100}},
  \bibinfo{pages}{030602} (\bibinfo{year}{2008}),
  \eprint{arXiv:cond-mat/0703314}.

\bibitem[{\citenamefont{Devi and Rajagopal}(2009)}]{Usha2009-thermo}
\bibinfo{author}{\bibfnamefont{A.~R.~U.} \bibnamefont{Devi}} \bibnamefont{and}
  \bibinfo{author}{\bibfnamefont{A.~K.} \bibnamefont{Rajagopal}},
  \bibinfo{journal}{Phys. Rev. E} \textbf{\bibinfo{volume}{80}},
  \bibinfo{pages}{011136} (\bibinfo{year}{2009}), \eprint{arXiv:0901.1453}.

\bibitem[{\citenamefont{Bloch et~al.}(2008)\citenamefont{Bloch, Dalibard, and
  Zwerger}}]{Bloch2008-thermo}
\bibinfo{author}{\bibfnamefont{I.}~\bibnamefont{Bloch}},
  \bibinfo{author}{\bibfnamefont{J.}~\bibnamefont{Dalibard}}, \bibnamefont{and}
  \bibinfo{author}{\bibfnamefont{W.}~\bibnamefont{Zwerger}},
  \bibinfo{journal}{Rev. Mod. Phys.} \textbf{\bibinfo{volume}{80}},
  \bibinfo{pages}{885} (\bibinfo{year}{2008}), \eprint{arXiv:0704.3011}.

\bibitem[{\citenamefont{Kinoshita et~al.}(2006)\citenamefont{Kinoshita, Wenger,
  and Weiss}}]{Kinoshita2006-thermo}
\bibinfo{author}{\bibfnamefont{T.}~\bibnamefont{Kinoshita}},
  \bibinfo{author}{\bibfnamefont{T.}~\bibnamefont{Wenger}}, \bibnamefont{and}
  \bibinfo{author}{\bibfnamefont{D.~S.} \bibnamefont{Weiss}},
  \bibinfo{journal}{Nature (London)} \textbf{\bibinfo{volume}{440}},
  \bibinfo{pages}{900} (\bibinfo{year}{2006}).

\bibitem[{\citenamefont{Hofferberth et~al.}(2007)\citenamefont{Hofferberth,
  Lesanovsky, Fischer, Schumm, and Schmiedmayer}}]{Hofferberth2007-thermo}
\bibinfo{author}{\bibfnamefont{S.}~\bibnamefont{Hofferberth}},
  \bibinfo{author}{\bibfnamefont{I.}~\bibnamefont{Lesanovsky}},
  \bibinfo{author}{\bibfnamefont{B.}~\bibnamefont{Fischer}},
  \bibinfo{author}{\bibfnamefont{T.}~\bibnamefont{Schumm}}, \bibnamefont{and}
  \bibinfo{author}{\bibfnamefont{J.}~\bibnamefont{Schmiedmayer}},
  \bibinfo{journal}{Nature} \textbf{\bibinfo{volume}{449}},
  \bibinfo{pages}{324} (\bibinfo{year}{2007}), \eprint{arXiv:0706.2259}.

\bibitem[{\citenamefont{{Short} and {Farrelly}}(2012)}]{Short2011-thermo}
\bibinfo{author}{\bibfnamefont{A.~J.} \bibnamefont{{Short}}} \bibnamefont{and}
  \bibinfo{author}{\bibfnamefont{T.~C.} \bibnamefont{{Farrelly}}},
  \bibinfo{journal}{New J. Phys.} \textbf{\bibinfo{volume}{14}},
  \bibinfo{pages}{013063} (\bibinfo{year}{2012}), \eprint{arXiv:1110.5759}.

\bibitem[{\citenamefont{Werner}(1989)}]{Werner-quantumstates}
\bibinfo{author}{\bibfnamefont{R.~F.} \bibnamefont{Werner}},
  \bibinfo{journal}{Phys. Rev. A} \textbf{\bibinfo{volume}{40}},
  \bibinfo{pages}{4277} (\bibinfo{year}{1989}).

\bibitem[{\citenamefont{Vinayak and Znidaric}(2012)}]{Vinayak2011-thermo}
\bibinfo{author}{\bibnamefont{Vinayak}} \bibnamefont{and}
  \bibinfo{author}{\bibfnamefont{M.}~\bibnamefont{Znidaric}},
  \bibinfo{journal}{J. Phys. A: Math. Theor.} \textbf{\bibinfo{volume}{45}},
  \bibinfo{pages}{125204} (\bibinfo{year}{2012}), \eprint{arXiv:1107.6035}.

\bibitem[{\citenamefont{Masanes et~al.}(2011)\citenamefont{Masanes, Roncaglia,
  and Acin}}]{Masanes2011-thermo}
\bibinfo{author}{\bibfnamefont{L.}~\bibnamefont{Masanes}},
  \bibinfo{author}{\bibfnamefont{A.~J.} \bibnamefont{Roncaglia}},
  \bibnamefont{and} \bibinfo{author}{\bibfnamefont{A.}~\bibnamefont{Acin}},
  \emph{\bibinfo{title}{The complexity of energy eigenstates as a mechanism for
  equilibration}} (\bibinfo{year}{2011}), \eprint{arXiv:1108.0374}.

\bibitem[{\citenamefont{{Cramer}}(2012)}]{Cramer2011-thermo}
\bibinfo{author}{\bibfnamefont{M.}~\bibnamefont{{Cramer}}},
  \bibinfo{journal}{New J. Phys.} \textbf{\bibinfo{volume}{14}},
  \bibinfo{pages}{053051} (\bibinfo{year}{2012}), \eprint{arXiv:1112.5295}.

\bibitem[{\citenamefont{Chatterjee}(2007)}]{Chatterjee2007-cm}
\bibinfo{author}{\bibfnamefont{S.}~\bibnamefont{Chatterjee}},
  \bibinfo{journal}{J. Funct. Anal.} \textbf{\bibinfo{volume}{245}},
  \bibinfo{pages}{379 } (\bibinfo{year}{2007}), \eprint{arXiv:math/0508518}.

\bibitem[{\citenamefont{Horn and Johnson}(1985)}]{Horn1985-book-mat}
\bibinfo{author}{\bibfnamefont{R.~A.} \bibnamefont{Horn}} \bibnamefont{and}
  \bibinfo{author}{\bibfnamefont{C.~R.} \bibnamefont{Johnson}},
  \emph{\bibinfo{title}{Matrix Analysis}} (\bibinfo{publisher}{Cambridge
  University Press}, \bibinfo{address}{Cambridge}, \bibinfo{year}{1985}).

\bibitem[{\citenamefont{{Popescu} et~al.}(2006)\citenamefont{{Popescu},
  {Short}, and {Winter}}}]{Popescu-thermo}
\bibinfo{author}{\bibfnamefont{S.}~\bibnamefont{{Popescu}}},
  \bibinfo{author}{\bibfnamefont{A.~J.} \bibnamefont{{Short}}},
  \bibnamefont{and} \bibinfo{author}{\bibfnamefont{A.}~\bibnamefont{{Winter}}},
  \bibinfo{journal}{Nat. Phys.} \textbf{\bibinfo{volume}{2}},
  \bibinfo{pages}{754} (\bibinfo{year}{2006}), \eprint{arXiv:quant-ph/0511225}.

\bibitem[{\citenamefont{{Brandino} et~al.}(2012)\citenamefont{{Brandino}, {De
  Luca}, {Konik}, and {Mussardo}}}]{Brandino2011-thermo}
\bibinfo{author}{\bibfnamefont{G.~P.} \bibnamefont{{Brandino}}},
  \bibinfo{author}{\bibfnamefont{A.}~\bibnamefont{{De Luca}}},
  \bibinfo{author}{\bibfnamefont{R.~M.} \bibnamefont{{Konik}}},
  \bibnamefont{and}
  \bibinfo{author}{\bibfnamefont{G.}~\bibnamefont{{Mussardo}}},
  \bibinfo{journal}{Phys. Rev. B} \textbf{\bibinfo{volume}{85}},
  \bibinfo{pages}{214435} (\bibinfo{year}{2012}), \eprint{arXiv:1111.6119}.

\bibitem[{\citenamefont{{Brand\~ao} et~al.}(2012)\citenamefont{{Brand\~ao},
  Harrow, and Horodecki}}]{Brandao-Harrow-Horodecki}
\bibinfo{author}{\bibfnamefont{F.~G. S.~L.} \bibnamefont{{Brand\~ao}}},
  \bibinfo{author}{\bibfnamefont{A.}~\bibnamefont{Harrow}}, \bibnamefont{and}
  \bibinfo{author}{\bibfnamefont{M.}~\bibnamefont{Horodecki}},
  \emph{\bibinfo{title}{Local random quantum circuits are approximate
  polynomial-designs}} (\bibinfo{year}{2012}), \eprint{arXiv:1208.0692}.

\bibitem[{\citenamefont{Fulton and J.Harris}(1991)}]{Fulton1991-book-rep}
\bibinfo{author}{\bibfnamefont{W.}~\bibnamefont{Fulton}} \bibnamefont{and}
  \bibinfo{author}{\bibnamefont{J.Harris}},
  \emph{\bibinfo{title}{Representation Theory - A first Course}}
  (\bibinfo{publisher}{Springer-Verlag}, \bibinfo{address}{New York},
  \bibinfo{year}{1991}).

\bibitem[{\citenamefont{Bannai and Ito}(1984)}]{Bannai1984-CAS}
\bibinfo{author}{\bibfnamefont{E.}~\bibnamefont{Bannai}} \bibnamefont{and}
  \bibinfo{author}{\bibfnamefont{T.}~\bibnamefont{Ito}},
  \emph{\bibinfo{title}{Algebraic combinatorics I.}}
  (\bibinfo{publisher}{Benjamin/Cumming Publishing Company},
  \bibinfo{year}{1984}).

\end{thebibliography}


\end{document}